# SymPlex Plots for Visualizing Properties in High-Dimensional Alloy Spaces

John Cavin*[1], Pravan Omprakash*[2], Adrien Couet[3], Rohan Mishra[1,2]

[1]Department of Mechanical Engineering & Materials Science, Washington University in St. Louis, St. Louis, MO, USA

[2]Institute of Materials Science & Engineering, Washington University in St. Louis, St. Louis, MO, USA

[3]Department of Nuclear Engineering & Engineering Physics, University of Wisconsin, Madison, WI, USA

*These authors contributed equally to this work.

Corresponding authors: J.C. (cavin@wustl.edu), R. M. (rmishra@wustl.edu)

**Keywords:** Multi-principal element alloys, data visualization, high entropy alloys, phase diagrams

**Abstract:** Conventional visualization tools such as phase diagrams and convex hulls are ill-suited to visualize multiple principal element alloys (MPEAs) due to their large compositional space that cannot be easily projected onto two dimensions. Here, SymPlex plots are introduced to enable the visualization of various properties along special paths in high-dimensional phase spaces of MPEAs. These are polar heatmaps that plot properties along high-symmetry paths radiating from the parent equimolar MPEA to a set of chosen lower-order compositions. SymPlex plots capture the changes in the energy landscape along the special paths and help visualize the effect of addition or substitution of components on the alloy stability, which can be especially useful to assess processing pathways for additive manufacturing. Thus, SymPlex plots can help guide design of MPEAs by showing connections between compositions and their properties in the high-dimensional phase space with more information concentrated near the equimolar region.

The discovery of high entropy alloys [1, 2], and more generally, multi-principal element alloys (MPEAs), has revitalized alloy design. Unlike most traditional alloys, which comprise one or two primary elements with other elements at lower concentrations, MPEAs consist of multiple principal elements in near-equimolar concentrations and expand the alloy phase space by opening up the unexplored central region. MPEAs can have a mixture of solid solutions and/or intermetallics [3]. The number of components in an MPEA can range from anywhere between three to over twenty [4, 5], offering a rich and complex design space to achieve a combination of desired properties such as toughness at cryogenic temperatures [6] or simultaneous strength and ductility [7]. Traditionally, isotherms and solidus/liquidus projections have been used to determine phase stability in binary and ternary systems. Extending these approaches to MPEAs necessitates more than three dimensions for visualization and thus becomes challenging. Beyond phase stability, assessing property variations with composition is also desirable to optimize alloys for targeted functionality. Thus, a scalable and intuitive visualization technique can help in effectively analyzing and designing high-dimensional MPEAs.

To address the dimensionality challenges in the visualization of MPEAs, several methods have been developed to reduce the plotting dimensions, each with specific strengths and limitations. Van de Walle et al. [8] approached the problem by reducing the number of plotting dimensions by one in a similar manner to representing quaternary systems with a stack of Gibbs-triangles. Phase boundaries in quinary alloy spaces were represented by arranging stacks of Gibbs-tetrahedra at fixed concentrations of the fifth member. More general techniques for projecting arbitrarily high-dimensional spaces into two dimensions (2D) such as affine projections and Uniform Manifold Approximation and Projection (UMAP) have been applied to visualize phase stability of MPEAs and their properties [9, 10]. An inevitable consequence of the projection of high-dimensional points into 2D is overcrowding, leading to a loss in interpretability. An entirely different strategy is to plot two relevant, non-compositional variables against each other and encode compositional information in labels instead. Inverse hull webs (IHW) [9] follow this approach by plotting two energy variables that effectively communicate thermodynamic phase stability and phase decomposition in a manner that scales to arbitrarily high dimensions. IHWs, however, do not provide information about general alloy properties. While all these methods have utility, there is not a method thus far that can unambiguously visualize properties within the high-dimensional composition space of MPEAs. A method that addresses these issues while focusing on visualizing the near-equimolar portion of a given composition space would be an especially beneficial addition to the suite of toolkits currently used to aid the design of MPEAs.

In this Letter, we present **Sym**metric-path Sim**plex** (SymPlex) plots as a method for visualizing high-dimensional phase spaces by focusing attention to high-symmetry paths radiating from the central equimolar composition to lower-order compositions. Data are encoded by color along radial arms extending from a single point corresponding to the equimolar composition. Composition-dependent properties can be plotted with SymPlex plots, making them useful tools for alloy design both for phase stabilization and property optimization. Because the paths displayed in SymPlex plots originate at the center of the phase space, the density of information is largest in the near-equimolar region where MPEAs exist. Tracing these paths from the equimolar point outward corresponds to an increase in concentration of some elements and a decrease of others. For a fixed total number of atoms, this can be viewed as substitution such as in the case of transmutation, but alternatively, it can be viewed as selective addition and deletion of elements or groups of elements. The effect on properties by the addition of components can be especially useful for additive manufacturing of MPEAs [11, 12]. Symplex plots can also aid in visualizing the effects of selective dissolution of elements during corrosion.

For a given alloy system, consisting of $N$ elements, an alloy of a specific composition can be defined by $N$ compositional variables $x_1, \ldots, x_N$, where the atomic fraction of element $i$ is $x_i$. Values of $x_i$ above 1 or below 0 are nonsensical and the sum of $x_i$ must be 1. These assertions can be expressed mathematically as:

$$0 \leq x_i \leq 1, \quad (1)$$

$$\sum_i^N x_i = 1. \quad (2)$$

For $N = 3$, Equations 1 and 2 define the barycentric coordinates of a regular unit triangle. For a general value of $N$, these barycentric coordinates correspond to an $(N-1)$-dimensional generalization of a triangle known as a regular unit $(N-1)$-simplex. Therefore, the composition space of a given $N$ component alloy family can be represented as the space enclosed by an $(N-1)$-simplex. Simplices have certain symmetries related to how the corner labels, in this case elements, permute. The group of symmetry operations that map an $(N-1)$-simplex on to itself is called $S_N$, the set of all permutations of element labels. The only point that is invariant under all actions of $S_N$ is the centroid of the simplex, corresponding to the equimolar alloy with each component having an atomic fraction of $1/N$.

Within a regular simplex, there are several special points and subsets corresponding to special alloys and alloy spaces. Fig. 1a depicts the compositional space of a quaternary alloy *ABCD* represented by a tetrahedron, also known as a 3-simplex. The surface of this tetrahedron consists of triangles (2-simplices), edges (1-simplices) and the vertices (0-simplices). This can be generalized to an $(N-1)$-simplex consisting of vertices, edges, faces, and hyperfaces of up to $(N-2)$ dimensions. Each of these subsystems are themselves lower dimensional simplices. In the case of the quaternary alloy depicted in Fig. 1a, there are 4 vertices representing the unary compositions, 6 edges representing the binary alloy systems, and 4 faces representing the ternary systems. For a general $N$-component alloy, the number of $(m-1)$-dimensional simplices on the surface corresponding to $m$-component alloy subsystems for $m = 1, \dots, N-1$ is ${}^{N}C_{m}$. The center of each of these simplices corresponds to the equimolar composition of the respective subsystem. We refer to the line segments drawn from the *N*-component equimolar alloy to the equimolar $m$-component subsystems as $m$-paths. These $m$-paths are invariant under a subset of $S_N$, $S_m \times S_{N-m}$, corresponding to mutually exclusive permutation within elements that are present ($S_m$) and not present ($S_{N-m}$) at the terminating composition. These line segments are high-symmetry axes of the simplex connecting the highest entropy compositions of the *N*-component alloy to the ${}^{N}C_{m}$ subsystems. The number of line segments is given by the binomial formula:

$$\sum_{m=1}^{N-1} {}^{N}C_{m} = 2^N - 2. \tag{3}$$

In Fig. 1a, there are a total of $2^4 - 2 = 14$ paths from the $N$-component equimolar alloy to lower-order equimolar $m$-component alloys. For every $m$-path, there is a corresponding antiparallel $(N-m)$-path that forms a continuous path connecting points on opposite sides of the $N$-simplex surface in a straight line through the centroid. Each dual pair of paths have the same symmetry, $S_m \times S_{N-m}$. For example, the line segment from *ABCD* to *ABC*, can be extended in the opposite direction to get the line segment drawn from *ABCD* to *D*. In addition to being invariant under $S_m \times S_{N-m}$, these $m$-paths have the added benefit of corresponding to elemental substitutions. For example, plotting along $(N-1)$-paths corresponds to the removal of one element from the $N$-component alloy, while keeping the other three in equal proportions. While a more general scheme for selecting paths through composition space is possible, as discussed below, from an alloy design and visualization standpoint, equimolar compositions serve as ideal reference points. They provide a chemically balanced baseline from which systematic substitutions can be explored.

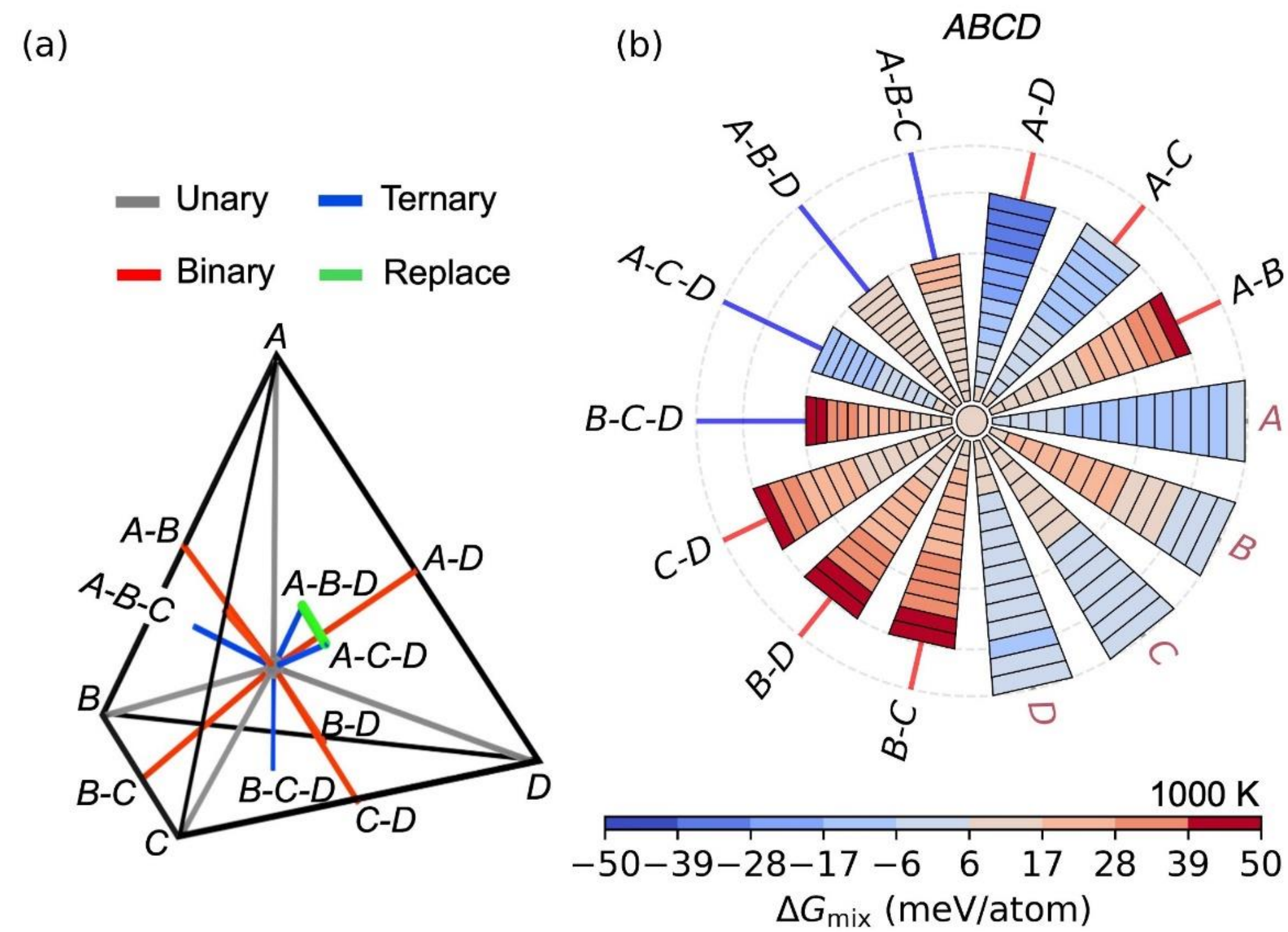

Fig. 1. (a) Visualization of a 4-component alloy space represented by a regular tetrahedron. Line segments from the 4-component equimolar alloy at the centroid or the barycenter to the equimolar $m$-order alloys represent various $m$-paths, colored by the value of $m$. An additional line from *ACD* to *ABD* represents replacing *C* with *B*. (b) Example SymPlex plot showing the Gibbs free energy of a regular solution toy-model of the *ABCD* alloy system.

The construction of a SymPlex plot involves selecting a subset of $m$-paths (typically including the complementary $(N - m)$ paths), a composition-dependent property, $X$, to plot, and a colormap to represent $X$. Fig. 1b shows an example of a SymPlex plot for the quaternary alloy system *ABCD*. The central circle corresponds to the equimolar quaternary alloy *ABCD*. Each arm of the plot extends radially outward from the center and corresponds to an $m$-path through the composition space, terminating on an $m$-component alloy. For each arm corresponding to an $m$-path with $m \leq N/2$, there is an arm on the opposite side corresponding to the complementary $(N - m)$-path. Together, these arm pairs correspond to continuous line segments that span the $(N - 1)$-dimensional simplex.

Another feature of the SymPlex plots is that distances are preserved from the composition space. The relative lengths of the arms in the SymPlex plots correspond to the relative lengths of the $m$-paths in a unit regular simplex with the absolute length normalized to the plot size. The length, $R$, of an $m$-path in an $(N - 1)$-dimensional unit regular simplex is given by:

$$R = \sqrt{\frac{(N-m)}{2mN}}. \tag{4}$$

The composition represented at each position along the arms can be parameterized by $t \in [0,1]$. $t = 0$ corresponds to the equimolar $N$-component alloy and $t = 1$ corresponds to the equimolar $m$-component alloy along the $m$-path. As stated earlier, the $m$-paths correspond to addition and deletion of components. The composition of all the components that increase along the $m$-path, can be written as a function of $t$:

$$x_i = \left(\frac{1}{m} - \frac{1}{N}\right) t + \frac{1}{N}, \tag{5}$$

and the components that decrease along the same $m$-path are given by:

$$x_d = \left(\frac{1}{N}\right)(1 - t). \tag{6}$$

Thus, SymPlex plots give an intuitive and quantitative sense of how ‘far’ a composition is away from the center of the phase diagram or the barycenter.

To demonstrate the utility of the SymPlex paradigm, we apply it to a model of Gibbs free energy of mixing using toy parameters. The Gibbs free energy of mixing of an alloy is given by

$$\Delta G_{mix} = \Delta H_{mix} - T\Delta S_{mix}, \tag{7}$$

where $\Delta H_{mix}$ is the enthalpy of mixing, $\Delta S_{mix}$ is the entropy of mixing, and $T$ is the temperature in absolute scale. The entropy of mixing may be modelled by the configurational entropy:

$$\Delta S_{config} = -k_B \sum_{i}^{N} x_i \ln x_i, \tag{8}$$

where $k_B$ is Boltzmann's constant. $\Delta H_{mix}$ may be modelled by a sum over pairwise enthalpy contributions using a regular solution model, which has been shown to work well for MPEA solid solutions [13]:

$$\Delta H_{mix} = \sum_{i=1}^{N} \sum_{j<i}^{N} \Omega_{ij} x_i x_j, \tag{9}$$

where $\Omega_{ij}$ are the interaction parameters for the constituent binary alloys fitted with regular solution models. Fig. 1b shows a SymPlex plot for the free energy of this system at 1000 K. For the toy model used in Fig. 1b, the values are 0.42 eV for $\Omega_{AB}$, $\Omega_{BC}$, $\Omega_{BD}$, and $\Omega_{CD}$, 0.21 eV for $\Omega_{AC}$, and 0.0105 eV for $\Omega_{AD}$.

For alloys up to 7 components, all $m$-paths can be shown with clarity; beyond this value, one must either select proper subsets or split the plots according to the order of the $m$-path. In Fig. 2a, only the unary-to-ternary paths are shown, the binary-to-binary paths have been omitted. Fig. 2b shows the complementary SymPlex plot of Fig. 2a with exclusively binary-to-binary paths. Splitting SymPlex plots allows for scaling to even high orders of alloys while maintaining legibility. Furthermore, omitting paths can leave room for extra information; in Fig. 2a, an additional path is shown between equimolar ternary alloys *ABD* and *ABC* corresponding to replacement of element *D* with *C*. This path, shown in green in Fig. 1b, is marked as path 1 in Fig. 2a. Similarly, Paths 2 and 3 are marked in Fig. 2a and Fig. 2b, respectively, and all three paths are plotted on a traditional *xy*-axis plot in Fig. 2c to help understand the function of the SymPlex plots.

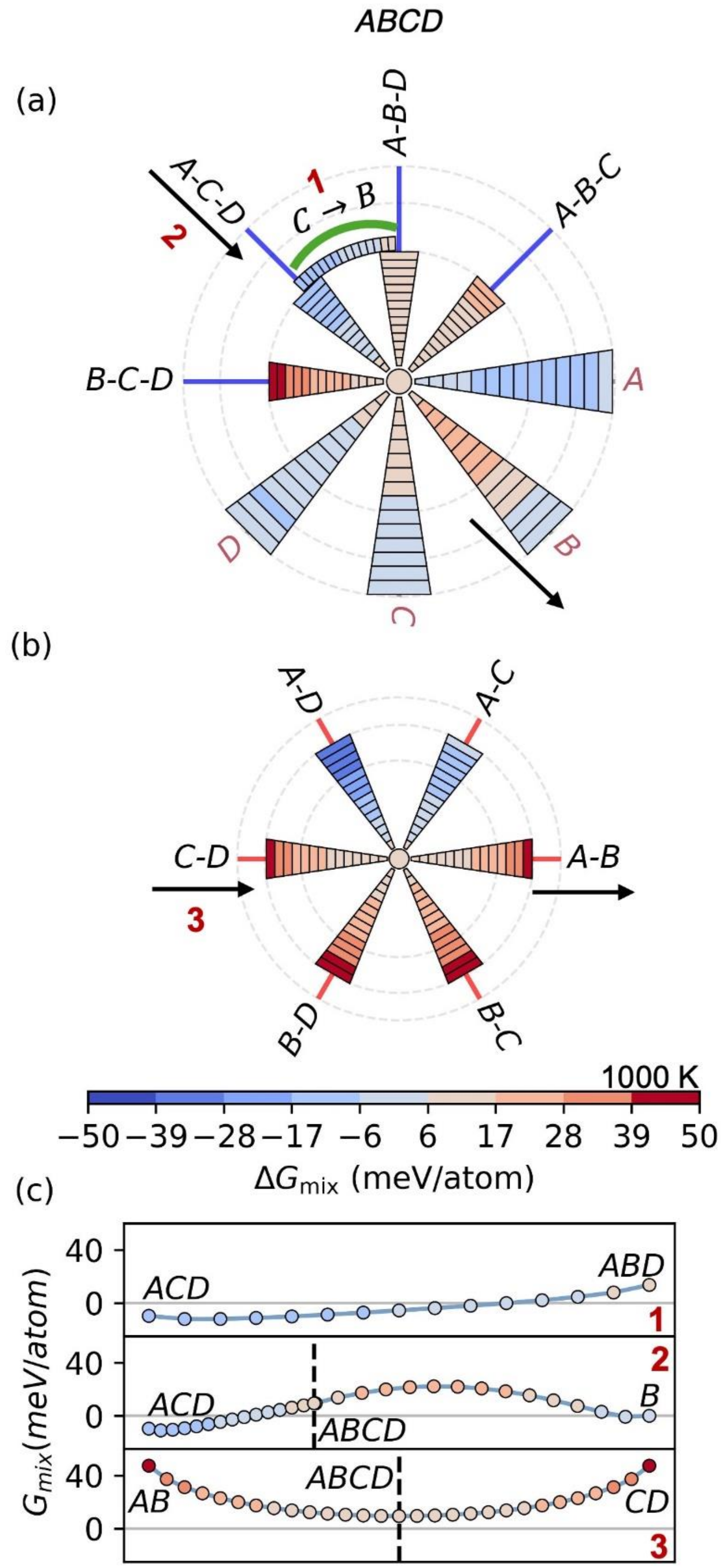
ABCD
(a)
A-B-D
A-B-C
A-C-D
1
C → B
2
B-C-D
A
B
C
D
(b)
A-D
A-C
C-D
A-B
3
B-D
B-C
1000 K
−50 −39 −28 −17 −6 6 17 28 39 50
ΔG_mix (meV/atom)
(c)
G_mix (meV/atom)
40
0
ACD
ABD
1
2
ACD
ABCD
B
ABCD
AB
CD
3

Fig. 2. (a) A reduced version of the SymPlex plot in Fig. 1a with 1-paths and 3-paths. An additional path between *ACD* and *ABD* is also shown. (b) The complementary SymPlex plot to Fig. 2a showing only 2-paths. (c) Gibbs free energy *vs*. composition along the paths 1, 2, 3 depicted in Figs. 2a and 2b.

Path 1, corresponding to the replacement of *C* with *B*, results in an increase in $\Delta G_{mix}$ from -9 to 13 meV/atom, shown by the coloration of Fig. 2a, thus implying that replacing the elements will result in phase separation. Path 2, corresponding to the path from *ACD* to *B* through *ABCD* shows a peak in free energy between *ABCD* and *B*. Such a path can be interpreted to be a miscibility gap between *ACD* and *B*. Path 3 in Fig. 2b, between the binaries *AB* and *CD*, shows a symmetrical trough, with *ABCD* having the minimum energy, as depicted in the lower panel of Fig. 2c. While *ABCD* may be locally stable against decomposition to *AB* and *CD*, the positive $\Delta G_{mix}$ suggests that phase segregation will still be preferred.

While plotting free energies can provide some qualitative understanding of stability, ultimately the ground-state phase decomposition is determined by the convex hull. To visualize stability using the SymPlex paradigm, we plot the energy above hull ($E_{\text{hull}}$), determined from a convex hull construction. $E_{\text{hull}}$ determines the stability of a specific composition, with a $E_{\text{hull}} = 0$ indicating a thermodynamically stable compound. Convex hulls were constructed using the methods employed by Evans et al. [9], utilizing their code and $\Delta H_{mix}$ values. Fig. 3a depicts the SymPlex plot for the quinary refractory system, Hf-Mo-Nb-Ti-Zr, at 300 K. The color bar was constructed to indicate compounds with a $E_{\text{hull}}$ of 0 $\text{meV/atom}$ in green (stable), a graded color scheme for metastable compounds with $1\ \text{meV/atom} \leq E_{\text{hull}} \leq 50\ \text{meV/atom}$ and unstable ones with $E_{\text{hull}} > 50\ \text{meV/atom}$ in dark red. First, it should be noticed that the equimolar quinary alloy is metastable. From a design perspective, looking at the regions near the central equimolar alloy, one can see which replacing elements will lead to stability closest to the equimolar condition: paths to Nb, Hf-Nb-Ti and Hf-Nb all look promising for stabilization at near equimolar values. Another trend that can be seen in Fig. 3a is the effect of individual elements on the quinary alloy. Higher fractions of Hf, Mo and Zr lead to immiscible regions, while Nb and Ti stabilize the alloy. The presence of combinations of these elements also dictates immiscibility in other regions of the phase diagram, such as the path from Hf-Mo-Nb-Ti-Zrto Hf-Mo-Zr is highly immiscible throughout.

The SymPlex plot in Fig. 3a, thus enables visualization of stability and metastability in specific regions of the phase diagram. The metastable and unstable regions of a solid solution alloy can be useful for two reasons. First, they can indicate spinodal decomposition and phase segregation if the system is known to not have many competing intermetallics. Second, a metastable solid solution can be potentially stabilized by increasing the temperature through the configurational entropy. Thus, the energy above hull of the solid solution can indicate the amount of temperature needed to stabilize it. However, SymPlex plots do not show decompositions of a metastable/unstable composition. For example, the equimolar quinary alloy, is found to be metastable from the convex hull analysis and decomposes to Hf-Zr and Mo-Nb-Ti. While Hf-Zr and Nb-Mo-Ti are shown to be stable in Fig. 3a, they are not easily interpreted to be the decomposition products for HfMoNbTiZr.

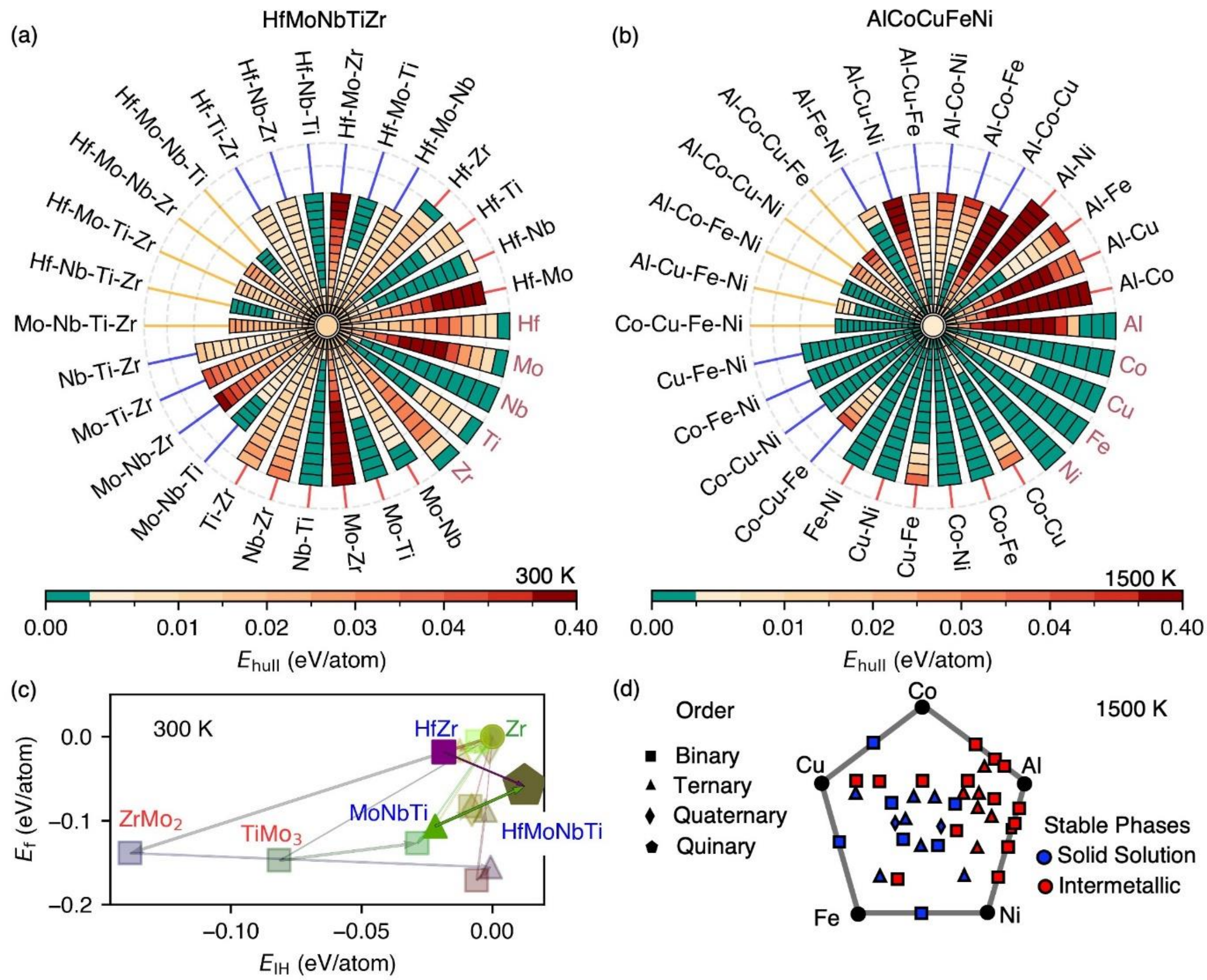

Fig. 3. (a) SymPlex plot showing the energy above hull ($E_{\text{hull}}$) for Hf-Mo-Nb-Ti-Zr system at 300 K. (b) SymPlex plot of $E_{\text{hull}}$ for Al-Co-Cu-Fe-Ni at 1500 K. (c) An Inverse Hull Web plot for the Hf-Mo-Nb-Ti-Zr system at 300 K. Blue and red text correspond to solid solutions and intermetallic phases, respectively. Circles, squares, triangles, diamonds, and pentagons correspond to compositions with 1, 2, 3, 4, and 5 components, respectively. Adapted from Evans et al. [9]. (d) Al-Co-Cu-Fe-Ni phase diagram at 1500 K projected onto 2 dimensions using barycentric coordinates. Blue and red markers indicate stable equimolar solid solutions and intermetallic phases, respectively. Adapted from Evans et al. [9] with permission.

To view the decomposition products readily, Inverse Hull Webs (IHW), proposed by Evans et al. [9], can be employed. IHWs forgo spatial compositional information altogether in favor of a purely energetic approach. They are plotted in 2 dimensions with the *y*-axis being the formation energy ($E_{\text{f}}$) and the *x*-axis being the inverse hull energy ($E_{\text{IH}}$), as shown in Fig. 3c. Positive values of $E_{\mathrm{IH}}$ are exactly the same as $E_{\mathrm{hull}}$, while a negative $E_{\mathrm{IH}}$ is obtained by removing the compound from the phase diagram, and recomputing the convex hull, and calculating the vertical distance between the compound's $E_f$ and the new hull. In Fig. 3c, the HfMoNbTiZr equimolar alloy (shown as a green pentagon) has a negative $E_f$ but has a positive $E_{\mathrm{IH}}$, indicating metastability. Arrows connecting the quinary equimolar to HfZr (purple square) and MoNbTi (green triangle), both having $E_{\mathrm{f}} < 0$ and $E_{\mathrm{IH}} < 0$, depict the decomposition products. Other information of the stable intermetallics in the alloy system (red labels), can also be read from the IHW. Hence, IHWs are useful to determine the decomposition products. In contrast, the Symplex plots depict contiguous regions of stability and metastability without directly displaying phase decomposition products. They can also help to identify processing pathways to stabilize the MPEA. Thus, IHWs and Symplex plots complement each other by representing different aspects of a high-dimensional convex hull. Together, these two plots enable the visualization of stable regions and directions, particularly near the higher-order equimolar alloy, and can guide synthesis efforts.

SymPlex plots can also be used to understand the effect of individual elements on the stability of a multinary solid solution. In Fig. 3b, a SymPlex plot of $E_{\mathrm{hull}}$for the Al-Co-Cu-Fe-Ni system at 1500 K is plotted. The *m*-paths are ordered in a way such that all the Al-containing compositions are plotted on the top half. The quinary equimolar solid solution AlCoCuFeNi is not stable, and any path that reduces the Al content stabilizes the solid solution. This is because, Al introduces stable intermetallics into the phase diagram, especially binary Ni-Al intermetallics as discussed below. The green regions in Fig. 3b represent stable solid solutions and are concentrated in the bottom half to highlight that Al destabilizes solid solutions. Furthermore, all the stable solid solutions in the top-half of Fig. 3b, shown in green color, contain both Al and Fe, which indicates that pairing Al with Fe can lead to stable solid solutions, as opposed to intermetallics. Even though the effect of Al on the stability of solid solutions can be seen clearly by the SymPlex plot of $E_{\mathrm{hull}}$, the stable intermetallics that contain Al cannot be determined.

For visualizing the stable phases present in a multinary alloy system, a 2D affine projection of the high-dimensional simplices can be used, as proposed by Evan *et al.* [9], and shown in Fig. 3d for the quinary Al-Co-Cu-Fe-Ni system at 1500 K. The vertices of the pentagon depict unary elements, while the edges of the pentagon represent binary systems between adjacent elements. Lines can be drawn within the pentagon, joining two vertices, for other binary systems. Triangles within the pentagon, connecting three vertices, depict ternary systems, with the centroid of the triangle corresponding to the equimolar ternary alloy. For quaternary systems, trapezoids can be drawn, and so on and so forth for higher dimensions. It can be seen that such a projection scheme can quickly become overcrowded and therefore allows plotting of only a limited number of compositions. The markers in Fig. 3d correspond to stable phases on the convex hull with red and blue markers denoting intermetallics and equimolar solid solutions, respectively. Stable, off-equimolar solid solutions are not shown for legibility. The presence of Al tends to cause the formation of intermetallic phases, as can be seen from the greater density of red markers near the Al corner of the pentagon. Many Al-Ni intermetallics appearing on the Al-Ni edge in Fig. 3d can be seen. So are Al-Cu intermetallics that sit in the interior of the diagram along with higher order alloys. Though these binary intermetallics and their specific compositions cannot be ascertained in the SymPlex plot in Fig. 3b, the effect of strongly competing intermetallics can be seen on the paths to Al-Cu, Al-Ni and Al-Co, which show energy above hulls greater than 0.4 eV/atom. Thus, SymPlex plots give a holistic view of the convex hull and depict the change in stability upon the inclusion of individual elements, or a pair of them.

Next, we discuss the use of SymPlex plots to show the variation in a chosen property with composition. One property of particular interest in the study of MPEAs is the configurational entropy, as defined by Eq. (8). In Fig. 4a, a SymPlex plot of configuration entropy of Cr-Mn-Fe-Co-Ni is plotted. The configurational entropy of an alloy is agnostic to the specific elements present; therefore, the elements chosen for Fig. 4a are merely placeholders. The highest order systems, in the center of the plot, have the highest magnitude of entropy, and the value decreases radially outward, becoming zero at the unary elements. Since relative distances in the compositional space along $m$-paths are conserved, it can be seen that entropy decreases faster along the $m$-paths to the unary points rather than to the binary points. This radially outward construction of the compositions in a SymPlex plots is intuitive to understand and can help visualize other composition-dependent properties such as elastic properties, density, and melting points.

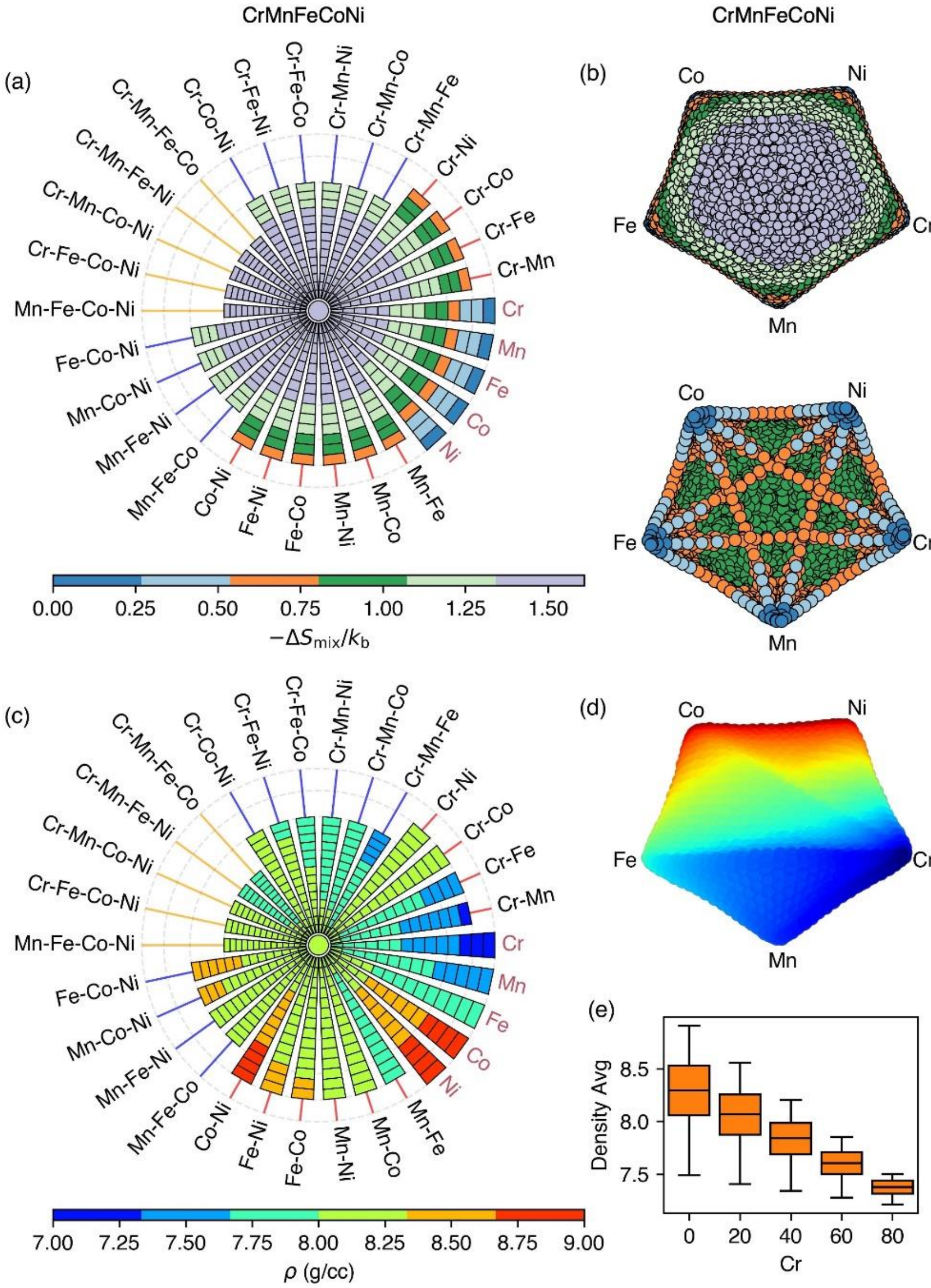

Fig. 4. (a) A SymPlex plot of configurational entropy for the Cr-Co-Fe-Mn-Ni alloy system (b) Configurational entropy of the same 5-component alloy system projected onto 2 dimensions using the UMAP paradigm viewed with ascending and descending values of the configurational entropy in the top and bottom panels, respectively. Adapted with permission from Vela et al. [10] . (c) A SymPlex plot showing the weight-averaged density of the Cr-Co-Fe-Mn-Ni alloy

system. (d) UMAP of the weight-averaged density of the Cr-Co-Fe-Mn-Ni alloy system with descending values. (e) A whisker-box plot showing the effect of Cr molar fraction on the density within the Cr-Co-Fe-Mn-Ni alloy system. Adapted with permission from Vela et al. [10].

Affine projections, like the one plotted in Fig. 3d, can be used to visualize entropy as well and supplement SymPlex plots. For such rule of mixing properties, affine projections qualitatively visualize the trend and also show the off-equimolar compositions not present in SymPlex plots and can hence complement them to visualize large areas of the phase space. In Fig. 4b, the affine projection is created by a dimensionality reduction method called UMAP [10]. Higher order compositions having higher values of entropy are located at the center of the pentagon, with the edges containing binary alloys. The compression of all the higher order compositions leads to overcrowding in these projections, requiring two views (top and bottom panel of Fig. 4b) corresponding to ascending and descending order of configurational entropy. Intermediate-order alloys, such as ternary and quaternary compositions, are unevenly distributed in UMAPs and often close to each other, leading to ambiguity. Therefore, they cannot be discerned from either the ascending or descending projections of entropy. Furthermore, complex properties of the phase space which cannot be obtained by averaging the composition/weight of the constituent elements like free energy, energy above hull, and order-disorder temperatures would be challenging to separate into ascending and descending schemes without losing clarity in the composition space. SymPlex plots are designed to overcome these challenges as discussed in Fig. 3 in the context of $E_{\text{hull}}$.

As another example of visualizing composition/order-dependent properties with a SymPlex plot, weighted-average of elemental density for the Cr-Co-Fe-Mn-Ni alloy system is shown in Fig. 4b. The compositions including Cr are selectively arranged on the top of the semicircle, to highlight the effect of Cr-addition on the density of these alloys. Yellow and red paths, having higher densities are constrained to the bottom semicircle, thus showing how Cr reduces the density of the alloys. It can also be seen that most parts of the central region of the system are green in color, with densities around 8 g/cc, not changing much with the addition/deletion of any element. The affine projection shown in Fig. 4d, captures the large variations in density between Co and Ni, and the other elements, but requires another box-and-whisker plot (shown in Fig. 4e) to understand the effect of increasing Cr content, specifically. SymPlex plots, thus, effectively condense both variations in trends across the composition space and the effect of addition/removal of specific elements. However, we note that the trends of the binaries and the ternaries within this subsystem are not shown in the SymPlex plots, the information presented here is solely related to the highest order alloys, and for more detailed information of the lower order alloys, similar SymPlex plots can be drawn or be visualized with affine projections [10].

So far, our focus has been on regions connected to the higher order equimolar composition, as these regions in the phase space offer a unique combination of high chemical disorder and lattice distortion, — that makes them attractive for select applications, such as enhanced performance in extreme environments . Equimolar and near-equimolar HEAs experience substantial lattice distortion and atomic mass variation [3]. These effects play a significant role in altering thermal and defect transport properties [2]. For example, one leading hypothesis regarding radiation resistance of near-equimolar HEAs is that atomic mass variation enhances phonon scattering, which reduces thermal conductivity and increases lifetime of thermal spikes during irradiation cascades [14]. A longer-lived heat spike promotes recombination of radiation-induced Frenkel pairs, thereby reducing the formation of stable point defects. Additionally, the complex chemical environment in an equimolar or near-equimolar HEA produces a highly variable migration energy landscape for point defects such as vacancies and interstitials. This variability would impede long-range defect migration and aggregation, suppressing the formation of voids and dislocation loops. These effects have been observed in equimolar CrFeMnNi-based alloys, where reduced swelling and radiation damage have been demonstrated under in-situ heavy-ion and dual-beam irradiation experiments [15].

However, we fully recognize that there is a vast and promising design space beyond equimolar compositions. Non-equimolar HEAs can be tuned to achieve targeted property profiles that may outperform equimolar counterparts in specific applications [16]. Recent research efforts have increasingly embraced high-throughput experimental and computational strategies to explore this broader compositional space [17]. By relaxing the symmetry conditions that originally motivated the creation of the SymPlex paradigm, a broader region of composition space can be explored using SymPlex-like plots. One such liberty that can be taken is centering the SymPlex plot around a non-equimolar composition. To illustrate the utility of this generalization, Figure 5a depicts $E_{\mathrm{hull}}$ with a binary colormap for the Al-Co-Cu-Fe-Ni system at 1500 K centered around $Al_7Cu_2Fe$, which is a stable intermetallic [18]. Note that this is the same system that is depicted in Figures 3b and 3d. The radiating paths in Fig. 5a end at the same lower order, equimolar compositions, as shown in Fig. 3b. Since all of the paths discussed in this Letter correspond to linear trajectories in the composition space, the elemental compositions will have forms similar to Eqs. 5 and 6 where some parameter $t$ will guide the trajectory from arbitrary start and end points along a straight line without necessarily having the symmetry of two groups of elements each defined by one shared atom fraction variable.

By examining these paths, one can make several observations on how to stabilize a solid-state solution. The direction towards the Al unary stabilizes a solid solution with the least amount of displacement in the composition space, resulting in a traditional Al-rich alloy with small solute concentrations. Moving towards Fe, results in stabilization of a solid solution that is closer to the ternary equimolar alloy of Al-Cu-Fe. Moving towards AlCoFeNi or CoCuFeNi result in quinary solid solutions, while moving towards any of the Al-containing ternary or binary alloys never escapes the formation of a mixture of the solid solution and an Al-rich intermetallic. This information can be used to probe the geometry of the hypervolume of mixed-phase region around a particular intermetallic phase. Fig. 5b illustrates this idea confined to the Al-Cu-Fe face of the composition space. The shaded region shows the interior of a convex hull of the phase boundaries shown in Fig. 5a.

An alternative expansion to the SymPlex method is to allow for paths to general points in the composition space. These paths may be allowed to violate permutation symmetry, and in fact, need not terminate on the boundaries of the composition space. Fig. 5c demonstrates these paths for the same Al-Co-Cu-Fe-Ni system but at 1800 K. The radiating arms point towards the unary corners of the composition space along with select stable intermetallic phases shown in Fig. 3d. Fig. 5c shows how “close” one can get in composition to these intermetallic phases before a solid solution, which is stable at the equimolar composition, is destabilized by the precipitation of intermetallic phases. Based on Fig. 5c, it appears that $Al_5Co_2$ and $Al_6Fe$ threaten the stability of near-equimolar solid solutions to a higher degree than $Al_2FeNi$. Another interesting feature is that unlike the other unary directions, Al initially destabilizes the solid solution phase by introducing more intermetallic phases, but then the solid solution phase becomes stable at higher Al concentrations corresponding to traditional Al-rich alloys. The selection by users of arbitrary central compositions and terminating arm compositions is a feature made available in the SymPlex package.

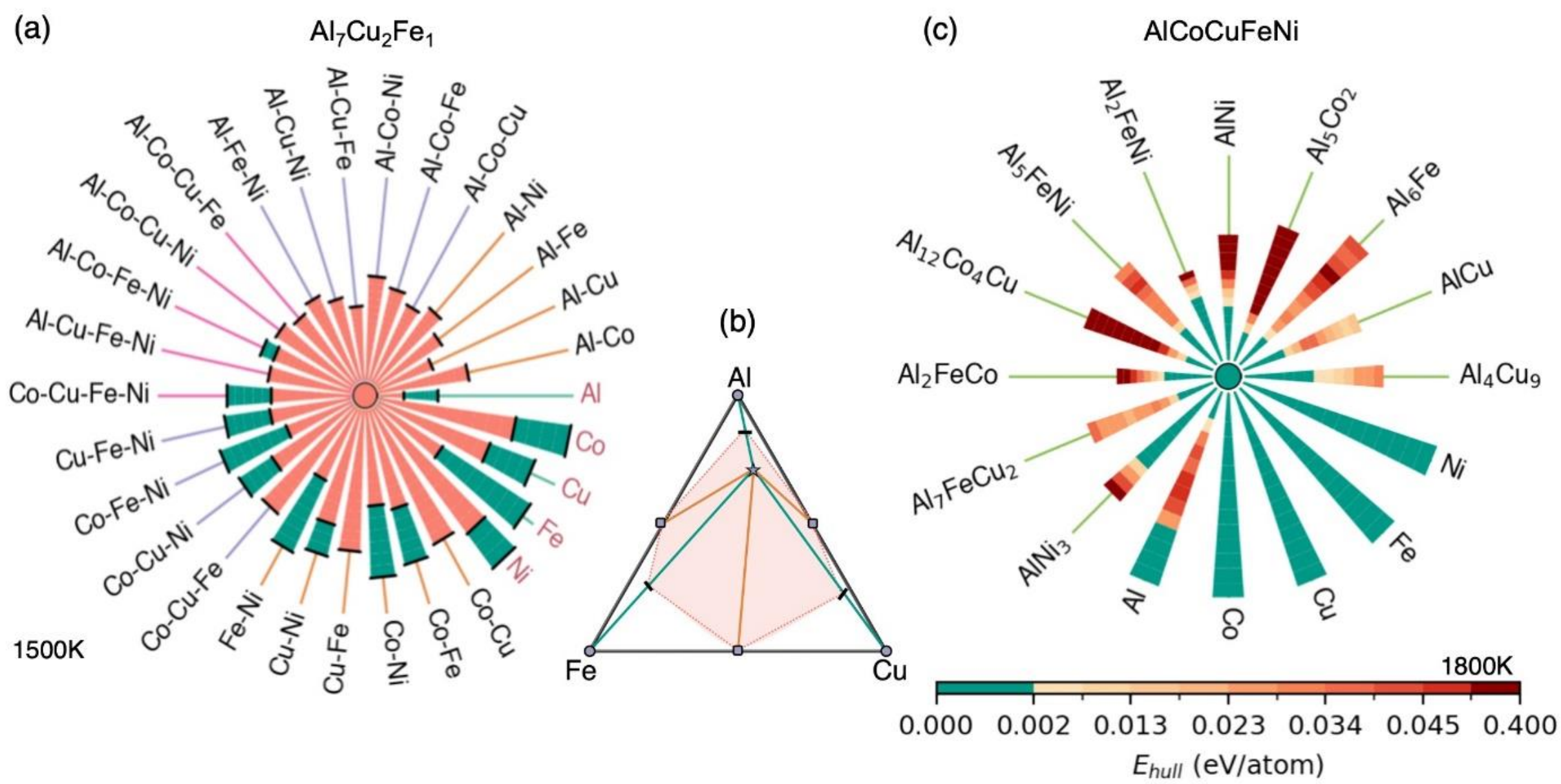

Fig. 5. (a) SymPlex plot of phase field where green indicates single-phase solid solution and pink indicates mixed phase for Al-Co-Cu-Fe-Ni at 1500 K, with the central point shifted from the quinary equimolar to a stable ternary intermetallic, $Al_{0.7}Cu_{0.2}Fe_{0.1}$, to depict the phase boundary between phase segregated and solid solution regions in the quinary phase space around the intermetallic (b) Al-Fe-Cu subspace at the boundary of the quinary Al-Co-Cu-Fe-Ni composition space. The shades region is enclosed by the convex hull of the phase boundaries marked in Fig. 5a as an approximation of the region where phase segregation occurs. (c) SymPlex plot of $E_{\mathrm{hull}}$ for Al-Co-Cu-Fe-Ni at 1800 K, with radial arms extending from the quinary equimolar to lower order stable intermetallics containing Al along with other unary end-members.

In summary, we have introduced SymPlex plots as an effective method for visualizing high-dimensional phase diagrams and compositional property trends in MPEAs. By leveraging the natural symmetry of simplices, SymPlex plots maintain one-to-one mapping with the high-dimension compositional space while maintaining high information density in the near-equimolar region. The paths chosen for SymPlex plots are particularly useful for mapping the properties associated with addition and removal of elements, and groups of elements. Thus, they can be used to represent intentional material manipulation such as additive manufacturing, and incidental processes like corrosion. We also show that SymPlex plots can be used to visualize arbitrary paths that do not end at or radiate from equimolar compositions, visualizing trends in other regions of the phase space, that can be chosen by the user based on the system and the application.

Furthermore, the ability to visualize arbitrary properties grants SymPlex plots the flexibility to be used for both stability analysis and property optimization. These factors make SymPlex plots a useful addition to the suite of visualization tools used for guiding MPEA design.

A Python code for constructing SymPlex plots is available at: https://github.com/Materials-Modelling-Microscopy/SymPlex. Example notebooks with detailed instructions for creating custom visualizations are also provided in the above GitHub repository.

**Acknowledgements**: This work was partially supported by the philanthropic initiatives of Eric and Wendy Schmidt through the Schmidt Family Foundation (J.C., A.C., R.M.) and the National Science Foundation through award # DMR-2145797 (P.O., R.M.). This work used computational resources through allocation DMR160007 from the Advanced Cyberinfrastructure Coordination Ecosystem: Services & Support (ACCESS) program, which is supported by NSF awards #2138259, #2138286, #2138307, #2137603, and #2138296.

## References

[1] B. Cantor, I.T. Chang, P. Knight, A. Vincent, Microstructural development in equiatomic multicomponent alloys, Materials Science and Engineering: A 375 (2004) 213-218.
[2] J.W. Yeh, S.K. Chen, S.J. Lin, J.Y. Gan, T.S. Chin, T.T. Shun, C.H. Tsau, S.Y. Chang, Nanostructured high-entropy alloys with multiple principal elements: novel alloy design concepts and outcomes, Advanced engineering materials 6(5) (2004) 299-303.
[3] D.B. Miracle, O.N. Senkov, A critical review of high entropy alloys and related concepts, Acta Materialia 122 (2017) 448-511.
[4] G. Cao, J. Liang, Z. Guo, K. Yang, G. Wang, H. Wang, X. Wan, Z. Li, Y. Bai, Y. Zhang, J. Liu, Y. Feng, Z. Zheng, C. Lu, G. He, Z. Xiong, Z. Liu, S. Chen, Y. Guo, M. Zeng, J. Lin, L. Fu, Liquid metal for high-entropy alloy nanoparticles synthesis, Nature 619(7968) (2023) 73-77.
[5] Y. Yao, Z. Huang, P. Xie, S.D. Lacey, R.J. Jacob, H. Xie, F. Chen, A. Nie, T. Pu, M. Rehwoldt, Carbothermal shock synthesis of high-entropy-alloy nanoparticles, Science 359(6383) (2018) 1489-1494.
[6] F. Otto, A. Dlouhý, C. Somsen, H. Bei, G. Eggeler, E.P. George, The influences of temperature and microstructure on the tensile properties of a CoCrFeMnNi high-entropy alloy, Acta Materialia 61(15) (2013) 5743-5755.
[7] Z. Li, K.G. Pradeep, Y. Deng, D. Raabe, C.C. Tasan, Metastable high-entropy dual-phase alloys overcome the strength–ductility trade-off, Nature 534(7606) (2016) 227-230.
[8] A. van de Walle, H. Chen, H. Liu, C. Nataraj, S. Samanta, S. Zhu, R. Arroyave, Interactive exploration of high-dimensional phase diagrams, JOM 74(9) (2022) 3478-3486.
[9] D. Evans, J. Chen, G. Bokas, W. Chen, G. Hautier, W. Sun, Visualizing temperature-dependent phase stability in high entropy alloys, npj Computational Materials 7(1) (2021) 151.
[10] B. Vela, T. Hastings, M. Allen, R. Arróyave, Visualizing high entropy alloy spaces: methods and best practices, Digital Discovery 4(1) (2025) 181-194.
[11] W. Zhang, A. Chabok, B.J. Kooi, Y. Pei, Additive manufactured high entropy alloys: A review of the microstructure and properties, Materials & Design 220 (2022).
[12] D.B. Miracle, M. Li, Z. Zhang, R. Mishra, K.M. Flores, Emerging Capabilities for the High-Throughput Characterization of Structural Materials, Annual Review of Materials Research 51(1) (2021) 131-164.
[13] Z.H. Zhang, M. Li, J. Cavin, K. Flores, R. Mishra, A Fast and Robust Method for Predicting the Phase Stability of Refractory Complex Concentrated Alloys using Pairwise Mixing Enthalpy, Acta Materialia 241 (2022) 118389.
[14] C.-L. Lu, S.-Y. Lu, J.-W. Yeh, W.-K. Hsu, Thermal expansion and enhanced heat transfer in high-entropy alloys, J. Appl. Crystallogr. 46 (2013) 736-739.
[15] C. Parkin, W.-Y. Chen, M. Li, K. Sridharan, A. Couet, Microstructural evolution of compositionally complex solid-solution alloys under in-situ dual-beam irradiation, Journal of Nuclear Materials 589 (2024) 154827.
[16] D.H. Cook, P. Kumar, M.I. Payne, C.H. Belcher, P. Borges, W. Wang, F. Walsh, Z. Li, A. Devaraj, M. Zhang, Kink bands promote exceptional fracture resistance in a NbTaTiHf refractory medium-entropy alloy, Science 384(6692) (2024) 178-184.
[17] A. Couet, Integrated high-throughput research in extreme environments targeted toward nuclear structural materials discovery, Journal of Nuclear Materials 559 (2022) 153425.

[18] A. Chemin, D. Marques, L. Bisanha, A.d.J. Motheo, W.W. Bose Filho, C.O.F. Ruchert, Influence of Al7Cu2Fe intermetallic particles on the localized corrosion of high strength aluminum alloys, Materials & Design 53 (2014) 118-123.